\documentclass{article}

\usepackage[english]{babel}
\usepackage[utf8x]{inputenc}
\usepackage[T1]{fontenc}
\usepackage[title]{appendix}
\usepackage{framed}
\usepackage{color}
\usepackage{algorithm}
\usepackage{algpseudocode}
\usepackage{float}
\usepackage{enumerate}
\usepackage{enumitem}
\usepackage[a4paper,top=3cm,bottom=2cm,left=3cm,right=3cm,marginparwidth=1.75cm]{geometry}
\usepackage{amsmath}
\usepackage{amssymb}
\usepackage{amsthm}
\usepackage{proof}
\usepackage{graphicx}
\usepackage{subcaption}
\usepackage[colorinlistoftodos]{todonotes}
\usepackage[colorlinks=true, allcolors=blue]{hyperref}
\usepackage{wrapfig,lipsum}
\usepackage{authblk}
\usepackage{multirow}
\usepackage{hhline}
\usepackage{setspace}
\usepackage{palatino}
\usepackage{mathdots}
\usepackage{multirow}
\usepackage{siunitx}
\usepackage{xcolor}
\usepackage{booktabs,colortbl, array}

\newcolumntype{C}[1]{>{\centering\arraybackslash}p{#1}}

\newtheorem{conjecture}{Conjecture}

\newtheorem{theorem}{Theorem}

\begin{document}

\title{Alibaba Cloud Quantum Development Platform: Applications to Quantum Algorithm Design}

\author[1]{Cupjin Huang\footnote{cupjin.huang@alibaba-inc.com}}
\author[1]{Mario Szegedy}
\author[1,2]{Fang Zhang}
\author[1]{Xun Gao}
\author[1]{Jianxin Chen} 
\author[1]{Yaoyun Shi}

\affil[1]{Alibaba Quantum Laboratory, Alibaba Group USA, Bellevue, WA 98004, USA}
\affil[2]{Department of Electrical Engineering and Computer Science, University of Michigan, Ann Arbor, MI 48109, USA}
\setcounter{Maxaffil}{0}
\renewcommand\Affilfont{\itshape\small}
    
\date{\today}
\maketitle
\begin{abstract}
    We report our work on the Alibaba Cloud Quantum Development Platform (AC-QDP). The capability of AC-QDP's computational engine was already reported in \cite{CZH+18, ZHN+19}. In this follow-up article, we  
    demonstrate with figures how AC-QDP helps in testing large-scale quantum 
    algorithms (currently within the QAOA framework). 
    We give new benchmark results on regular graphs. AC-QDP's QAOA framework can simulate thousands of qubits for up to $4$ layers. Then we discuss two interesting use cases we 
    have implemented on the platform: 1. Optimal QAOA sequences for small-cycle free graphs; 2. Graph structure discovery.
\end{abstract}

\medskip

\section{Introduction}

In the past two years significant efforts have been made towards building intermediate-scale quantum devices (NISQs). These devices are quantum circuits that
execute computation tasks before decoherence kicks in, so that we can forget about the daunting task of quantum error correction.
Assume we could achieve the goal of building devices with a few hundreds high-quality qubits.
The next task is then to find conservatively constructed clever quantum algorithms, likely with the help of classical simulation.
Unfortunately, even the above modest parameter regime forbids the idea of classically storing the entire state of the quantum machine, due to 
exploding memory requirements above more than roughly fifty qubits. 

The Alibaba Cloud Quantum Development 
Platform (AC-QDP) aims to utilize Alibaba's massive classical computational resources for aiding the development of quantum 
applications and quantum computers themselves,
with a design, which in some cases, although not always, defeats the ''curse of dimensionality.'' Albeit our tensor-network based idea 
is not new, and a few other platforms have adopted similar principles \cite{SSC+17}, the implementation details allow us to
reach farther limits than before.

The capability of AC-QDP's computational engine was already reported in \cite{CZH+18, ZHN+19}. In this follow-up article, we  
focus on showing how AC-QDP helps in testing of large-scale quantum 
algorithms, currently within the QAOA framework.
First we report on new benchmark results, then we discuss two interesting use cases we 
have implemented on the platform: 1. Optimal QAOA sequences for small-cycle free graphs; 2. Graph structure discovery.

\section{An extensible platform for QAOA research}

Quantum Approximate Optimization Algorithm (QAOA) was developed by Farhi, Goldstone and Gutman
to adapt Quantum Adiabatic algorithms to the quantum circuit model  \cite{farhi2014quantum}. However, it is not a straightforward adaptation, but rather a one with a twist.
QAOA, depending on the number of its {\em layers}, cannot fully emulate Quantum Adiabatic algorithms. In compensation, it offers 
a new feature: a QAOA circuit has free parameters that are to be optimized 
in a feed-back process, making it a prime example to a hybrid quantum-classical algorithm.

\subsection{Quantum Approximate Optimization Algorithm (QAOA)}
Many combinatorial optimization problems, including MAX-CUT,
MAX-$k$-SAT, Maximum Independent Set, Quadratic Unconstrained Binary Optimization (QUBO), can be formulated as finding the minimum of a function $C:\{0,1\}^n\rightarrow \mathbb{R}$, which can be decomposed as a sum of local terms $C=\sum_{i=1}^m C_i$ each acting only on a small number of bits. We call this class of optimization problems Weighted Constraint Satisfaction Problem (WCSP). Although this class is NP/NPO-hard, numerous approximation algorithms and heuristics exist for finding near-optimal solutions in theory and practice.

The Quantum Approximate Optimization Problem (QAOA) tries to solve such optimization problems in a quantum variational approach: Regarding the objective function $C:2^n\rightarrow \mathbb{R}$ as a local Hamiltonian $\hat{C}=\sum_x f(x)|x\rangle\langle x|=\sum_{j=1}^m\hat{C}_j$, QAOA takes the ansatz that the state
\begin{equation}\label{annealingQAOA}
|\vec{\gamma}, \vec{\beta}\rangle = e^{-i\beta_{p} \hat{B}} e^{-i\gamma_{p} \hat{C}} \cdots e^{-i\beta_{1} \hat{B}} e^{-i\gamma_{1} \hat{C}} (|+\rangle)^{\otimes n}
\end{equation}
defined by the mixing operator $\hat{B}=\sum_{i=1}^nX_i$ and the angle sequences $\vec{\gamma}, \vec{\beta}\in \mathbb{R}^d$ approaches the ground state of $\hat{C}$ with carefully chosen parameters $\vec{\gamma}, \vec{\beta}$, even with a small number of layers $p$. In the case that the objective function $C$ takes integer values, it is sufficient to restrict to $\vec{\gamma}, \vec{\beta}\in[0, 2\pi]^d$. Since 
$$e^{-i\beta \hat{B}}=\bigotimes_{j=1}^ne^{-i\beta X_j}$$
can be decomposed into individual $X$-rotations, and
$$e^{-i\gamma \hat{C}}=\prod_{j=1}^me^{-i\gamma \hat{C}_j}$$
can be decomposed into commuting unitary operations each acting on a small number of qubits, a state $|\vec{\gamma}, \vec{\beta} \rangle$ can be readily prepared using a quantum circuit.

The QAOA consists of two phases; in the first phase, (near) optimal angle sequences $(\vec{\gamma}, \vec{\beta})$ are obtained; in the second phase, these parameters are used to generate a certain number of copies of the state $|\vec{\gamma},\vec{\beta}\rangle$ via a quantum circuit. These states are then measured under the computational basis to yield a classical string $z$ such that $C(z)$ is close to the minimum value of $C$.

The QAOA energy function with $p$ layers is defined as follows:
$$F_p(\vec{\gamma},\vec{\beta}):=\langle \vec{\gamma},\vec{\beta}|\hat{C}|\vec{\gamma},\vec{\beta}\rangle.$$
The operational meaning of $F_p(\vec{\gamma},\vec{\beta})$ is the expectation value of $C(Z)$, where the random string $Z$ comes from measuring the quantum state $|\vec{\gamma},\vec{\beta}\rangle$ under the computational basis. The QAOA energy function indicates the performance of the QAOA algorithm itself, and is commonly used as the objective function for finding good angle sequences $\vec{\gamma},\vec{\beta}$ via a classical optimizer.

\subsection{Classical simulation of QAOA}
The two most basic simulation approaches for QAOA are state-vector simulations and tensor-network-based simulations.

Given the angle sequences $\vec{\gamma},\vec{\beta}$, state-vector simulations simulate the entire process of generating the state $|\vec{\gamma},\vec{\beta}\rangle$ and computing the energy function. The functionality of sampling from the QAOA distribution comes automatically with such simulations, but those simulations quickly grow infeasible when the size of the quantum system increases. 

The tensor-network-based approach focuses on evaluating the QAOA energy function by regarding it as the sum of energy contributions from each Hamiltonian term $\langle \vec{\gamma},\vec{\beta}|\hat{C}_j|\vec{\gamma},\vec{\beta}\rangle$. By regarding each such energy term as a separate tensor network, one can use tensor-network contractions to evaluate them individually. Such an approach is sometimes much more efficient than the state-vector simulation, whereas sampling becomes a non-trivial task since the state-vector information may not be present at any point. In the case that only the QAOA energy function value is needed, tensor-network based approach allows us to handle much bigger instances than we could ever hope for with the state-vector approach. The tensor-network-based simulation for the QAOA has already been implemented in~\cite{SSC+17}, but our software was able to solve QAOA instances with both larger sizes and larger depths, due to improvements in our implementations that we discuss in Section~\ref{sec:imp}.

\section{Implementation and Benchmarking}
\label{sec:imp}
\subsection{Implementation}
Based on our existing tensor network library, we have implemented a QAOA simulation package inside AC-QDP. The main structure of the software is as follows:
\paragraph{Preprocessing.} Altough tensor network contraction can in principle be much more efficient than state-vector evolution, determining the optimal contraction order can be itself time-consuming, and is in general NP-hard~\cite{MS08}. In the scenario of QAOA, tensor network contraction is done many times, whereas the structure of the tensor networks corresponding to a particular hamiltonian term is invariant with respect to the angle sequences. This allows us to determine the contraction orders ahead of querying in the preprocessing phase to reduce the time consumption of each energy query.

The following techniques help further reduce the complexity of the tensor networks. We would like to point out that the following techniques have been observed by others~\cite{BIS+17,farhi2014quantum} and have been applied in other softwares as well~\cite{SSC+17,BIS+17}. The advantage of our software does not come from a single technique, rather our implementation of the software that successfully incorporates all of the below.
\begin{description}
    \item[Lightcones] We apply the technique, first observed in~\cite{farhi2014quantum}, to cancel out circuit components that commute through the local hamiltonian term. In other words, only qubits and gates that lie in the \emph{lightcone} of the Hamiltonian term need to be taken into account. This greatly simplifies the tensor networks, especially in cases whre the Hamiltonian is sparse and the depth of the QAOA is very small.
    \item[Diagonality] We further simplify the tensor networks by making use of the fact that all gates of the form $e^{-i\gamma \hat{C}_j}$, including the Hamiltonian terms, are diagonal. Similar to what we have done in~\cite{ZHN+19}, (block) diagonal operations in quantum circuits can be more concisely expressed in terms of tensors, and the simplified tensor networks usually result in much shorter contraction time.
    \item[Uniform superposition] Since the starting state of QAOA consists of only plus states, one further simplification of the tensor networks is to replace the tensors representing the plus states by summation of the corresponding index, with proper normalization.
\end{description}

Upon receiving a WCSP instance, the software constructs each individual QAOA lightcone as a tensor network template. The optimal contraction orders are then determined by tree decomposition softwares. There is also the option of switching to ordinary state-vector evolution if it appears to be more favorable.
\paragraph{Querying.} Upon receiving an angle sequence, the software puts into each of the tensor network templates the tensors determined by the angle sequence. Tensor network contractions are then performed according to the predetermined orders, in the previously built tensot entwork contraction library.
\paragraph{Optimization.} The software incorporates many existing global and local classical optimizers. To optimize, a classical optimizer will be running on top of the QAOA energy querying method, indicating which angle sequences to be queried next. When the classical optimizer finishes, the expected energy, together with the near-optimal angle sequence, will be output by the software.

\subsection{Benchmarking Results}

To compare our QAOA simulator with existing quantum simulation package equipped with QAOA functionalities, we choose MAX-CUT problems on a random regular graph and compare the time spent for a single energy function query. Given a undirected graph $G=(V,E)$, we assume without loss of generality that $V=[n]$. The cut function is then
$$C(z)=\sum_{(i,j)\in E}w_{ij}(z_i\oplus z_j).$$
Here we consider the case where the graph is unweighted, i.e. the case where $w_{ij}=1$ for all $(i,j)\in E.$

We tested our software with three of the mainstream quantum computing softwares: Cirq~\cite{Cirq}, Qiskit~\cite{Qiskit} and qTorch~\cite{SSC+17} on an Alibaba Cloud ECS instance with $24$ Intel Xeon Gold 6149 @ 3.1 GHz and $96$ gigabytes of memory. Cirq and Qiskit use the state-vector evolution, whereas qTorch is based on tensor-network simulation with a featured QAOA implementation.

Each software is given the same random instance of a regular graph, and is asked to perform 5 times the energy query to obtain an average time for a single query. The regular graphs are chosen randomly, running through 10, 20, 30, 50, 100 and 1000 vertices with degree 3, 4 and 5. Fig.~\ref{fig:bench} shows the performance of AC-QDP compared to other softwares, for degree $3$ cases:

\begin{figure}[hbpt]
    \centering
    \includegraphics[width=\textwidth]{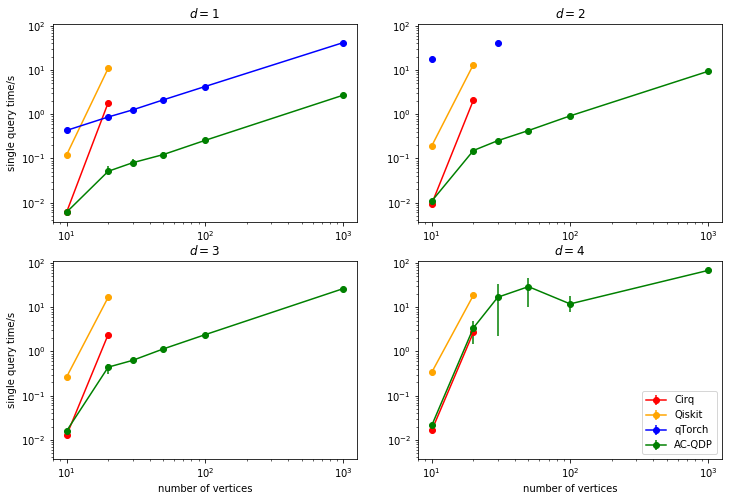}
    \caption{Benchmarking results for random $3$-regular graphs. For each number of vertices, $5$ random instances are drawn, and QAOA of depth 1 through 4 are simulated. The average time for each single energy query is recorded in the above figure. For the case $d=2$, qTorch succeeded $1$ out of $5$ times when $n=10$, and $1$ out of $5$ times when $n=30$. The corresponding data points are from the cases where it succeeds.}
    \label{fig:bench}
\end{figure}

From Figure~\ref{fig:bench}, it can be seen that the AC-QDP simulator is able to handle instances where the number of qubits is very large, while maintaining performance comparable to mainstream state-vector-based simulators when the number of qubits is within the range that can be handled by state-vector simulations. We notice a large variation in the running time for the cases where $d=3,p=4$ and the number of vertices $n=30$ or $50$. For these cases, the local structures from each edge are much more diverse, resulting in lightcones with very different contraction complexities. Such variation is supressed when the number of vertices further increases, since small cycles are less likely to appear and the local structures tends to be more tree-like.

We also demonstrate the capability of our software regarding very large regular graphs. For regular 3-, 4- and 5- graphs with 1000 vertices, we are able to perform a single query efficiently, with depth up to 5. We report the running time for a single query in Table~\ref{tbl:1000}.
\begin{table}[h]
    \centering
\begin{tabular}{|c|ccc|}
    \hline
    & $d=3$ & $d=4$ & $d=5$ \\
    \hline
    $p=1$ & 2.707 & 4.318 & 6.789\\
    $p=2$ & 9.433 & 20.455 & 36.913\\
    $p=3$ & 26.450 & 80.816& *\\
    $p=4$ & 68.651 & *& *\\
    $p=5$ & 409.022 & *& *\\
    \hline
\end{tabular}
\caption{Time for a single query in seconds, for regular graphs with $n=1000$ vertices.}
\label{tbl:1000}
\end{table}

\section{Optimization for small-cycle-free regular graphs}
We apply our QAOA software on a family of instances called small-cycle-free regular graphs. When a regular graph has girth at least $2p+2$, the lightcones originating from any edge would only contain vertices in the tree-like neighbor of that edge with distance $p$. In other words, all $k$-regular graphs with the same number of vertices and a girth at least $2p+2$ yeild identical QAOA energy functions $F_p$, and the simulation of QAOA on such instances can be readily reduced to the single lightcone on the unique tree-like subgraph, parametrized by the degree $d$ and the depth $p$.

These small-cycle-free instances are interesting for the following reasons:
\begin{itemize}
    \item Local approximation algorithms for max-cut of graphs have also been studied in the literature, both for classical algorithms~\cite{bamas2019local} and for quantum algorithms~\cite{hastings2019classical}. In~\cite{hastings2019classical}, 1-level QAOA is compared with classical one-step local algorithm on triangle-free regular graphs. It is shown that the scaling of the performance of QAOA is a factor worse than certain simple classical local algorithms.
    \item A large random regular graph is almost always small-cycle free. This suggests that the optimal angle sequences for the tree-like case is usually a good starting point for the QAOA when dealing with large regular graphs. In fact, it is proposed in~\cite{streif2019training} that the angle sequences for small-cycle-free instances be used directly to completely get rid of the optimization phase, which could be chanllenging on noisy quantum devices.
\end{itemize}

Existing works have analytically solved the QAOA values for cases where $k=2$ or $p=1$~\cite{wang2018quantum}. It has also been pointed out that analytically solving for the cases where $d\geq 2$ could be chanllenging. By incorporating several optimization techniques, we were able to solve for the scores of for these cases to the best of our knowledge.

\begin{table}[ht]
    \centering
    \begin{tabular}{cc|c|c|ccccc}
        $d$ & $p$ &\#\ vertices&time p.q.[s]& \texttt{dlib} & \texttt{de} & \texttt{FOURIER} & \texttt{procession} & grid search\\
        \hline
        3&1& 6&0.0015&  \textbf{0.692}&  \textbf{0.692}&   \textbf{0.692}&   \textbf{0.692}&   \textbf{0.692}\\
        3&2& 14&0.0048&  0.738&  \textbf{0.756}&   \textbf{0.756}&   \textbf{0.756}&   \textbf{0.756}\\
        3&3& 30& 0.0129& 0.782&  0.782&   0.782&   \textbf{0.792}&   \textbf{0.792}\\
        3&4& 62& 0.0349& \textbf{0.817}&  0.805&   0.811&   0.806&     -\\
        3&5& 126&0.0945& \textbf{0.822}&    -&   0.819&   0.800&     -\\
        \hline
        4&1&8& 0.0019&  \textbf{0.662}&  \textbf{0.662}&   \textbf{0.662}&   \textbf{0.662}&   \textbf{0.662}\\
        4&2&26& 0.0076&  \textbf{0.716}&  \textbf{0.716}&   \textbf{0.716}&   \textbf{0.716}&   \textbf{0.716}\\
        4&3&80& 0.0306&  0.737&  0.737&   0.739&   0.735&   \textbf{0.749}\\
        4&4&242& 0.1432&  0.760&  \textbf{0.761}&   0.751&   0.753&     -\\
        \hline
        5&1&10& 0.0021&  \textbf{0.643}&  \textbf{0.643}&   \textbf{0.643}&   0.357&   \textbf{0.643}\\
        5&2& 42& 0.0113& \textbf{0.691}&  \textbf{0.691}&   \textbf{0.691}&   0.682&   \textbf{0.691}\\
        5&3& 170& 0.0666& 0.709&  0.709&   \textbf{0.720}&   0.711&   \textbf{0.720}\\
        5&4&682& 0.6436&  0.731&    -&   0.736&   \textbf{0.739}&     -\\
        \hline
        6&1& 12& 0.0025& \textbf{0.629}&  \textbf{0.629}&   \textbf{0.629}&   \textbf{0.629}&   \textbf{0.629}\\
        6&2&62& 0.0160&  \textbf{0.673}&  \textbf{0.673}&   \textbf{0.673}&   \textbf{0.673}&   \textbf{0.673}\\
        6&3& 312& 0.1422& 0.690&  0.690&   \textbf{0.699}&   0.692&   \textbf{0.699}\\
        \hline
        7&1& 14& 0.0027& \textbf{0.619}&  \textbf{0.619}&   \textbf{0.619}&   \textbf{0.619}&   \textbf{0.619}\\
        7&2&86& 0.0218&  \textbf{0.659}&  \textbf{0.659}&   \textbf{0.659}&   \textbf{0.659}&   \textbf{0.659}\\
        7&3&518&  0.2663& \textbf{0.674}&    -&   0.667&   0.667&     -\\
        \hline
    \end{tabular}
    \caption{Comparison of different optimizers on small-cycle-free regular graphs, together with the best values obtained. Cases where the experiment was infeasible is indicated by `-'. Best function values found are highlighted. The number of vertices and the averaged time per query in AC-QDP are also listed.}
    \label{tbl:optimal}
\end{table}

Table~\ref{tbl:optimal} is a comparison of several optimization strategies and their performance on optimizing the small-cycle-free regular graphs. For global optimization techniques, we used \texttt{find\_min\_global} from the software package \texttt{dlib} and differential evolution method from \texttt{scipy}, indicated as \texttt{de} in the following table. We also tried the FOURIER heuristic from~\cite{zhou2018quantum}. A similar heuristic is proposed in~\cite{streif2019training}. It starts from training the best pair of angles for the case $p=1$ via local search, and use the angles as the starting point of the local search for the case $p=2$. In each iteration, the number of layers is increased by $1$ and the previously solved angle sequences are used as the starting point for the new local search. We call this heuristic \texttt{procession} in this work. Finally, we performed a grid search for cases where with depth up to 3 on the Alibaba Cloud ECS cluster. The most promising starting points are then fed into local optimizers to yield the final result.

Thanks to the tensor-network-based implementation, our software is able to efficiently query small-cycle-free instances of large degree $d$ and number of layers $p$, albeit the number of vertices involved in the tree-like subgraphs might be large. However, with the increase of the number of angles, optimization . We show in Table~\ref{tbl:optimal} the optimization results for the following cases: $d=3$ and $p\leq 5$, $d=4,5$ and $p\leq 4$, and $d=6,7$ and $p=3$.

We also report here the best angle sequences that we found in Table~\ref{tbl:t}. Note that many angle sequences lead to the same result due to the symmetry of the problem; the angle sequences we report here are just the particular choices found by our optimizers.
\begin{table}[ht]
    \begin{subtable}{\linewidth}\centering
    {\begin{tabular}{c|ccccc}
        $p$&1&2&3&4&5\\
        \hline
        $\gamma_1,\beta_1$&2.5261,1.1781&2.4488,0.6541&2.7197,0.9619&0.4088,5.6836&3.8370,1.6617\\
        $\gamma_2,\beta_2$&-&0.1450,1.3665&5.4848,2.6820&0.7806,1.1365&2.0107,0.5446\\
        $\gamma_3,\beta_3$&-&-&2.2046,1.8064&0.9880,5.9864&5.2067,1.0590\\
        $\gamma_4,\beta_4$&-&-&-&4.2985,4.8714&0.2183,4.0065\\
        $\gamma_5,\beta_5$&-&-&-&-&1.7976,6.0988\\
    \end{tabular}}
    \caption{Angle sequences for $d=3$.}
\end{subtable}
\vspace{0.5cm}
\begin{subtable}{\linewidth}\centering
    {\begin{tabular}{c|cccc}
        $p$&1&2&3&4\\
        \hline
        $\gamma_1,\beta_1$&2.6180,0.3926&2.1446,0.3768&0.3545,0.9829&0.3288,0.9793\\
        $\gamma_2,\beta_2$&-&0.9934,0.4666&3.7929,2.7184&3.9103,2.0612\\
        $\gamma_3,\beta_3$&-&-&3.8958,2.9186&6.0978,2.2043\\
        $\gamma_4,\beta_4$&-&-&-&4.0656,2.9514\\
    \end{tabular}}

    \caption{Angle sequences for $d=4$.}
\end{subtable}
\vspace{0.5cm}
\begin{subtable}{\linewidth}\centering
    {\begin{tabular}{c|cccc}
        $p$&1&2&3&4\\
        \hline
        $\gamma_1,\beta_1$&3.6052,0.3926&3.5003,0.5241&2.8300,0.9925&0.3210,1.0004\\
        $\gamma_2,\beta_2$&-&3.7866,2.8628&2.5780,1.9784&3.6348,0.3995\\
        $\gamma_3,\beta_3$&-&-&2.4938,1.3518&3.6786,2.8340\\
        $\gamma_4,\beta_4$&-&-&-&0.6376,1.4060\\
    \end{tabular}}
    \caption{Angle sequences for $d=5$.}
\end{subtable}
\vspace{0.5cm}
\begin{subtable}{\linewidth}\centering
    {\begin{tabular}{c|ccc}
        $p$&1&2&3\\
        \hline
        $\gamma_1,\beta_1$&5.8624,0.3926&2.8173,2.0890&2.8602,2.1434\\
        $\gamma_2,\beta_2$&-&2.5619,0.2765&5.7791,0.3989\\
        $\gamma_3,\beta_3$&-&-&2.5648,1.7877\\
    \end{tabular}}
    \caption{Angle sequences for $d=6$.}
\end{subtable}
\vspace{0.5cm}
\begin{subtable}{\linewidth}\centering
    {
    \begin{tabular}{c|ccc}
        $p$&1&2&3\\
        \hline
        $\gamma_1,\beta_1$&3.5292,0.3926&3.4398,0.5144&3.4002,2.1397\\
        $\gamma_2,\beta_2$&-&0.5312,0.2751&0.4604,0.3933\\
        $\gamma_3,\beta_3$&-&-&3.6669,1.3551\\
    \end{tabular}}
    \caption{Angle sequences for $d=7$.}
\end{subtable}
    \caption{Best angle sequences found by classical optimizers in AC-QDP.}
    \label{tbl:t}
\end{table}

\section{Graph Structure Discovery}

In this section we apply QAOA towards discovering similarity in structures of graphs.
Graph Isomorphism is the strongest 
notion of similarity:
$G_{1}$ and $G_{2}$, are isomorphic if there is a 
1-1 edge-preserving map between the
vertex sets of $G_{1}$ and $G_{2}$. Two isomorphic graphs always give the same QAOA energies under the same
fixed degree sequence $(\vec{\beta},\vec{\gamma})$  (the optimization step does not play a role here).
One can potentially discover non-isomorphism with the algorithm:

\begin{enumerate}
\item Pick $(\vec{\gamma},\vec{\beta}) \in [0,2\pi]^{2p}$ uniformly at random, for predetermined number of layers $p$.
\item Compute  $E_{1} = F^{(1)}_p(\vec{\gamma},\vec{\beta})$ and $E_{2} = F^{(2)}_p (\vec{\gamma},\vec{\beta})$, where $F^{(1)}_p$ and $F^{(2)}_p$
are the QAOA energy functions associated with the CUT function of the two graphs.
\item Output ``yes'' if $E_{1}=E_{2}$, otherwise output ``no.'' (Repeat the procedure for larger certainty.)
\end{enumerate}

\begin{conjecture}[\cite{marioXx4}]
With probability 1 over all degree sequences in $[0,2\pi]^{2d}$ the above algorithm gives different energy values when $G_{1}$ and $G_{2}$
on $n$ nodes are non-isomorphic.
\end{conjecture}

Even when the conjecture holds, it is not sufficient to put the graph isomorphism problem in quantum polynomial time, since
the difference between the energy values can be exponentially small.

Graph isomorphism with classical resources is a true success story. In 1980 Babai and Erd\H{o}s have shown that for dense random graphs already the naive classification algorithm 
arrives at a perfect refinement of the node set  \cite{DBLP:journals/siamcomp/BabaiES80}. Luks in 1982 gave a beautiful 
polynomial time algorithm for bounded degree graphs \cite{DBLP:journals/jcss/Luks82}.
Finally, a breakthrough result of L\'aszl\'o Babai, obtained in 2016, ensures that graph isomorphism is in ${\rm exp}(C(\log n)^{c})$, 
i.e. in quasi polynomial classical time \cite{DBLP:conf/stoc/Babai16}. 

There have been several attempts to construct quantum operators from walks to identify isomorphism classes of graphs.
A promising recent attempt is that of
Kamil Br\'adler, Shmuel Friedland, Josh Izaac, Nathan Killoran and Daiqin Su  of Xanudu.
The authors have found out
that their boson sampling experiments always distinguishes between two non-isomorphic graphs \cite{1810.10644}.
Unfortunately, the argument that proves this fact does not give an efficient bound on the energy gap.

We have extensively tested the QAOA-based algorithm described above
with the Alibaba Cloud Quantum Development Platform, and we could
separate all non-isomorphic three regular graphs up-to size 18, all strongly regular graphs up-to size 26
and Miyazaki graphs of size 20. 
These findings and the theoretical results in \cite{marioXx4} make us believe 
that QAOA energies give a full characterization of isomorphism classes, unlike many quantum walk based distinguishers that 
were considered earlier \cite{2015JPhA48k5302W}, 
\cite{DBLP:journals/pr/EmmsSWH09} and \cite{Mahasinghe_2015}.
The table below gives pairs or classes of graphs and indicates the number of layers of the QAOA circuit that distinguishes between them :

\medskip

\begin{center}
\begin{tabular}{|c|c|} \hline
Class or Pair of Graphs & QAOA Depth Giving Full Separation  \\\hline\hline\\\\[-3.5\medskipamount]
Miyazaki I and II, 20 nodes & 4  \\\hline\\\\[-3.5\medskipamount]
Praust I and II, 20 nodes  & 4  \\\hline\\\\[-3.5\medskipamount]
All 4060 non-iso 3-regular graphs &  4  \\\hline\\\\[-3.5\medskipamount]
All 41301 non-iso 3-regular graphs on 18 nodes &  4  \\\hline\\[-1.5\medskipamount]
All 10 non-iso graphs in the SRG 26,10,3,4 family &  3  \\\hline
\end{tabular}
\end{center}

Our simulator could not cope with Cai-F\"urer-Immerman graphs of size 40 (a well-known pair of hard instances). 
The reason is that although we could compute the QAOA values of these graphs for depth 4, this depth was not enough to 
distinguish between the two non-isomorphic instances. Depth 5 for these graphs seems to be out of the scope of our simulator,
and we raise it as a challenging benchmark for other simulators.

\medskip

\noindent{\bf Similarity of Graphs.} There are several good algorithms that find isomorphism of graphs, so 
we have been seeking to deploy our QAOA based method in related areas. Discovering similarities between graphs with QAOA is a natural idea. 
We expect that the the difference in the QAOA energies of graphs
reflects the extent to which they are non-isomorphic.
Although it can occur that for a fixed degree sequence the QAOA energies of two very different graphs nearly (or completely) coincide by chance, 
when we select two or more sequences, the likelihood that for all of them this happens quickly goes to zero.
For a set $\{G_{1}, G_{2},\ldots\}$ of graphs on $n$ nodes, if we pick a small parameter $k$, for random degree
sequences $\vec{\beta}^{1},\vec{\gamma}^{1}, \ldots, \vec{\beta}^{k},\vec{\gamma}^{k}$, 
we can define a ``landscape'' by
representing $G_{i}$ with the vector
\[
x_{i} = [F_i(\vec{\beta}^{1},\vec{\gamma}^{1}), \ldots, F_i(\vec{\beta}^{k},\vec{\gamma}^{k})]
\]

\begin{figure*}[h!]\
    \centering
    \begin{subfigure}[b]{0.35\textwidth}
        \centering
        \includegraphics[width=\textwidth]{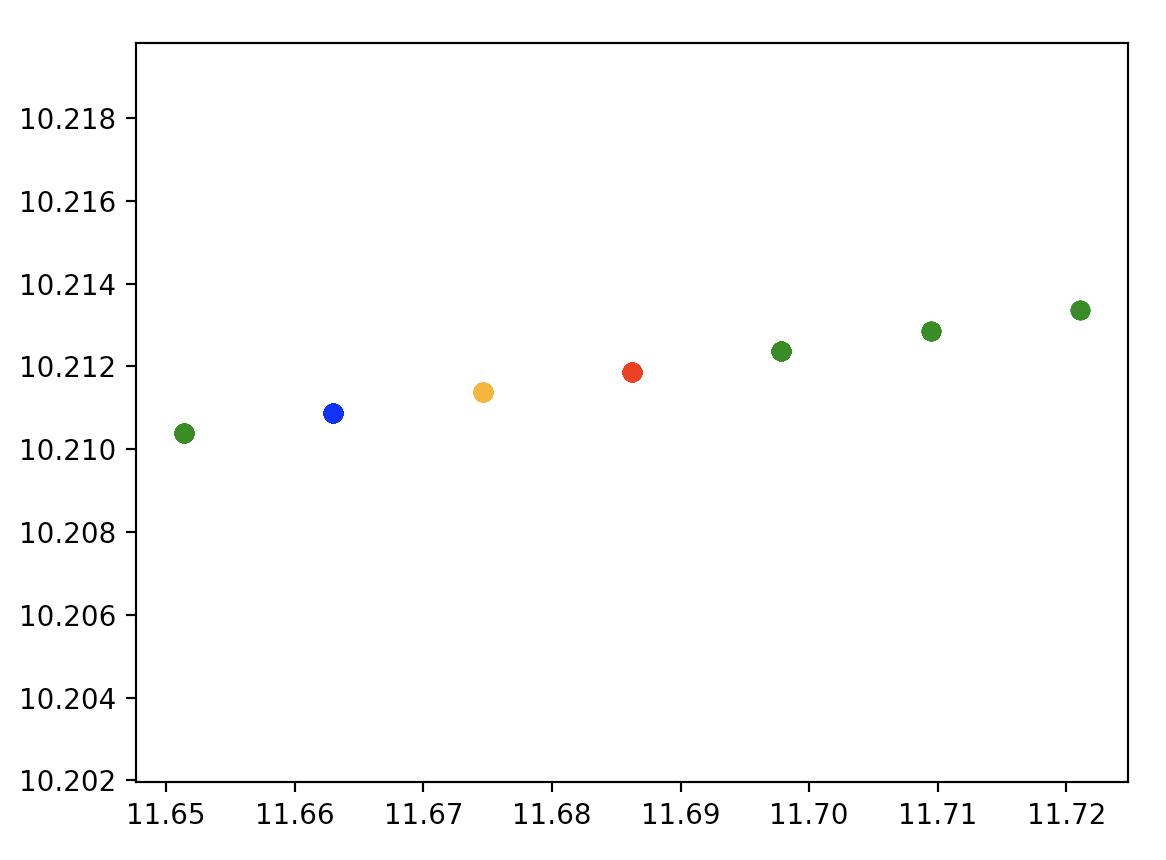}
        \caption{3-reg. graphs of size 14, $p=1$}
    \end{subfigure}
    \hspace{0.4in}
    \begin{subfigure}[b]{0.35\textwidth}
        \centering
        \includegraphics[width=\textwidth]{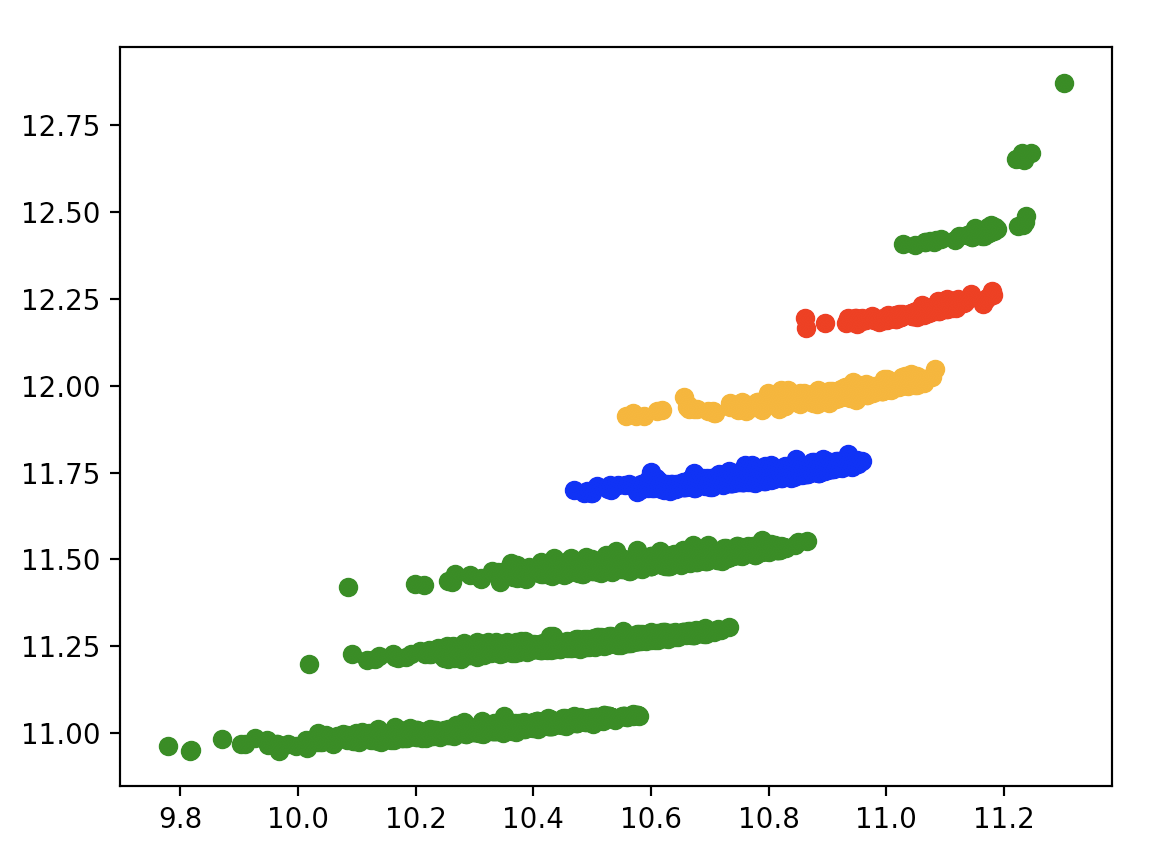}
        \caption{3-reg. graphs of size 16, $p=3$}
    \end{subfigure}%
   \\
    \begin{subfigure}[b]{0.35\textwidth}
        \centering
        \includegraphics[width=\textwidth]{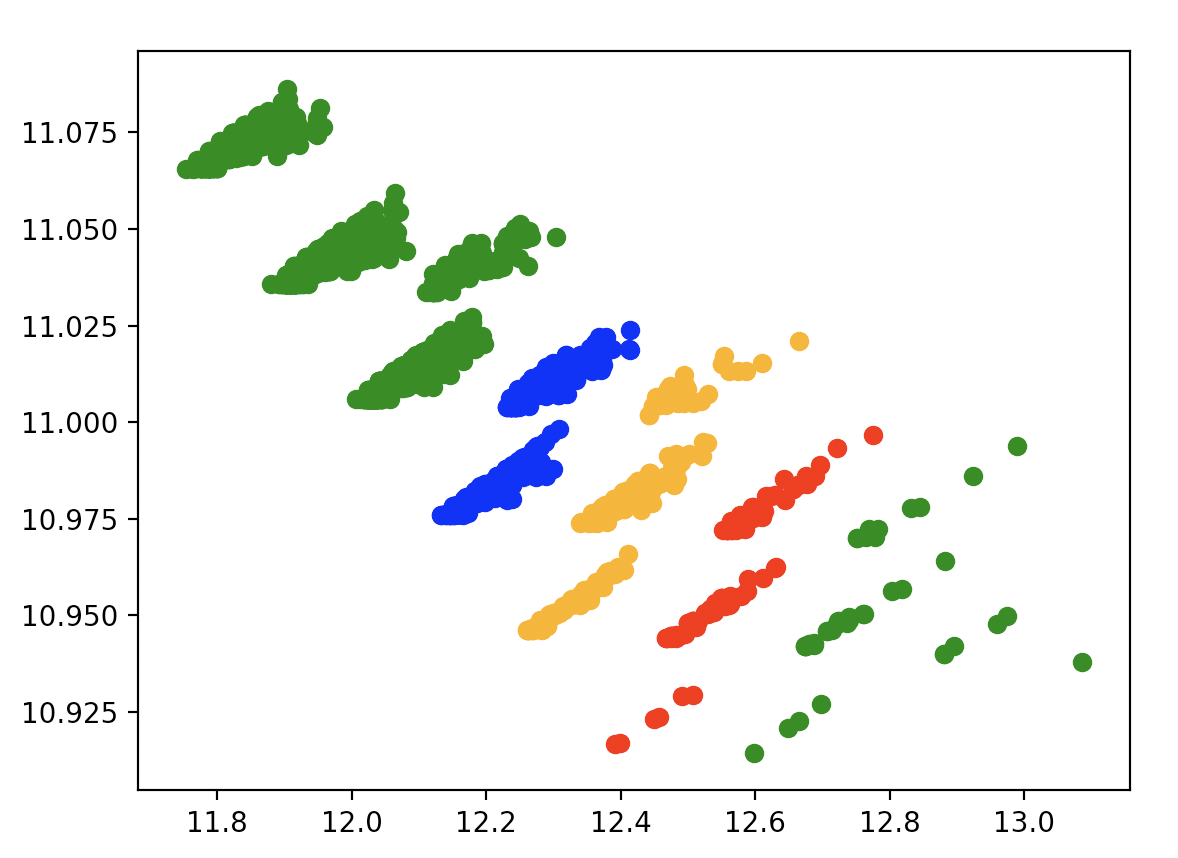}
        \caption{3-reg. graphs of size 16, $p=3$}
    \end{subfigure} 
    \hspace{0.4in}
    \begin{subfigure}[b]{0.35\textwidth}
        \centering
        \includegraphics[width=\textwidth]{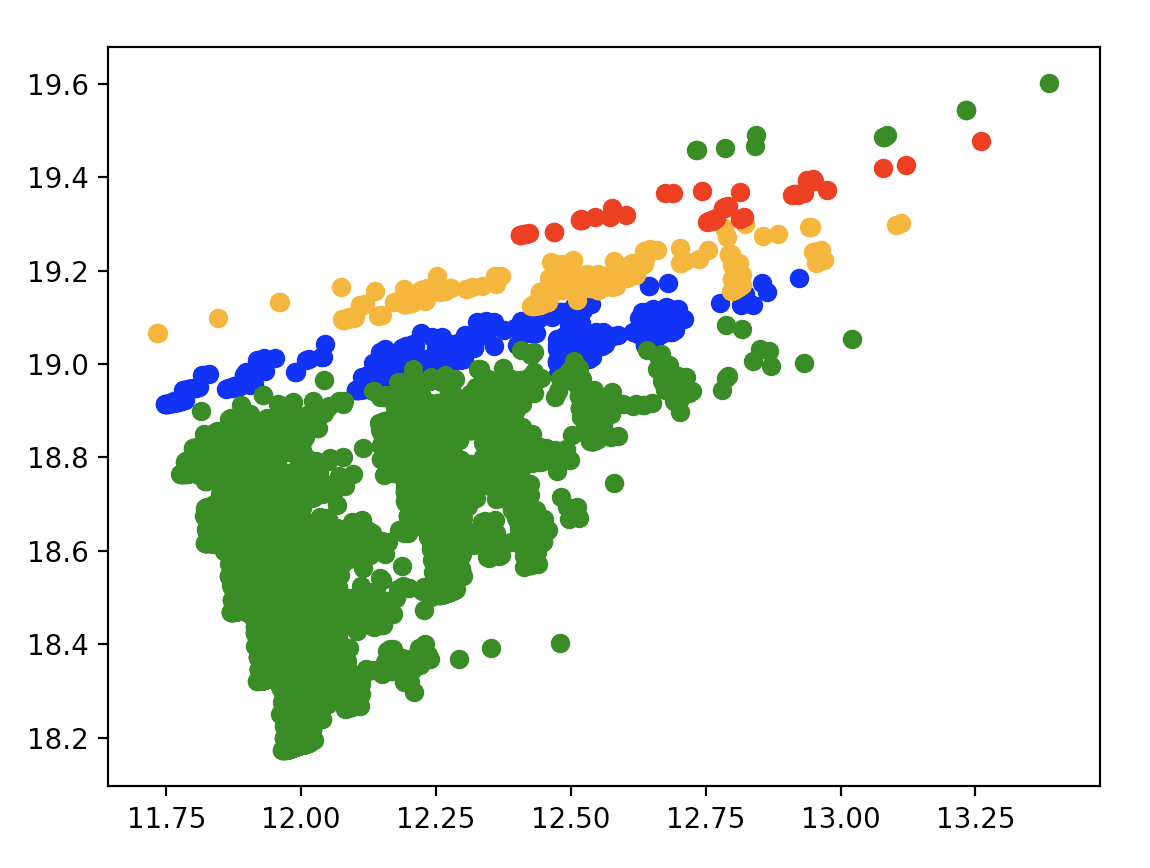}
        \caption{3-reg. graphs of size 18, $p=4$}
    \end{subfigure}%
    \caption{Different Landscapes of all 3-regular graphs on 16 and 18 nodes. \label{landscapes}
    Plots (b) and (c) are made from the same sets of graphs but using different angle sequences.}
\end{figure*}

\begin{figure*}[h!]\
    \centering
    \begin{subfigure}[b]{0.35\textwidth}
        \centering
        \includegraphics[width=\textwidth]{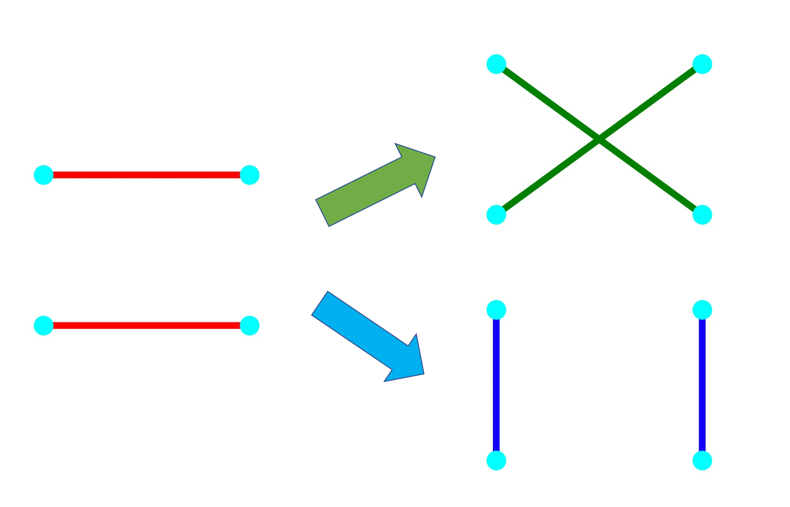}
        \caption{A walk step}
    \end{subfigure}
    \hspace{0.4in}
    \begin{subfigure}[b]{0.35\textwidth}
        \centering
        \includegraphics[width=\textwidth]{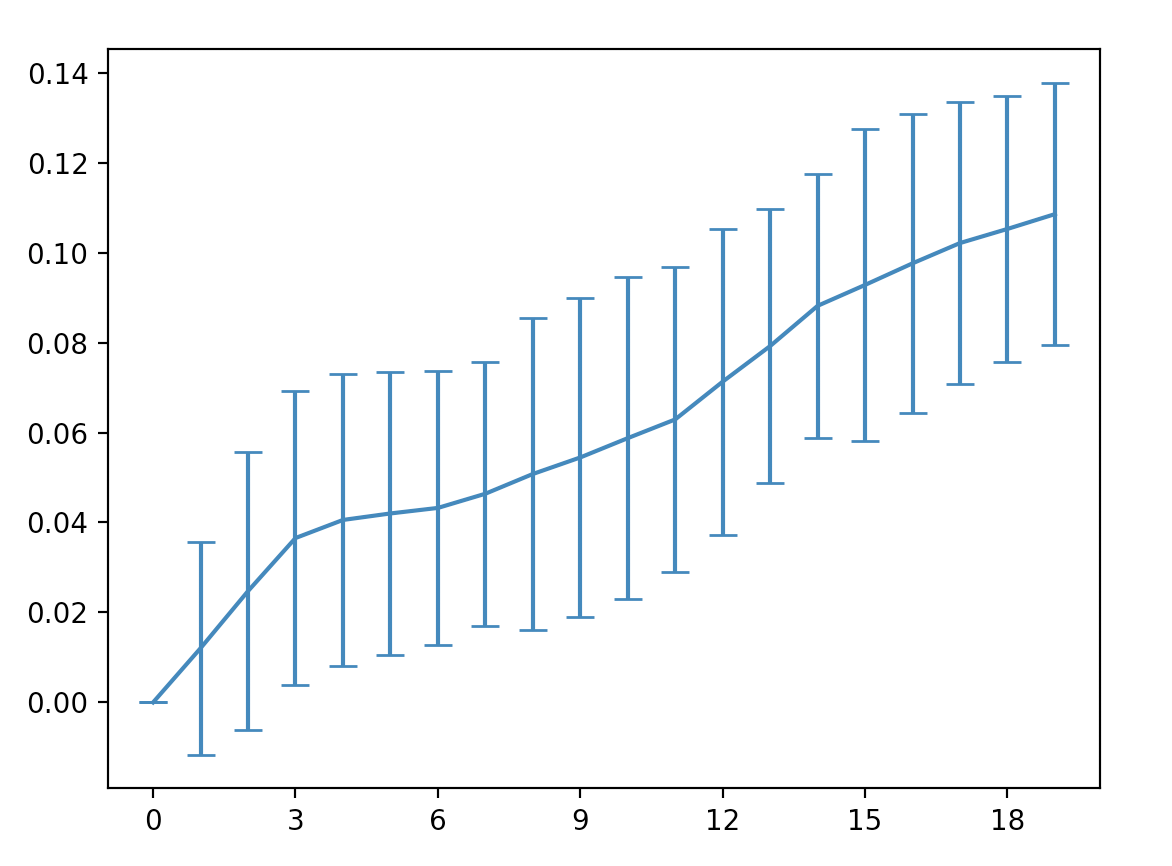}
        \caption{Walk on 5-regular graphs on 20 nodes, $p=3$; Increasing distance means larger QAOA separation}
    \end{subfigure}%
   \\
    \begin{subfigure}[b]{0.35\textwidth}
        \centering
        \includegraphics[width=\textwidth]{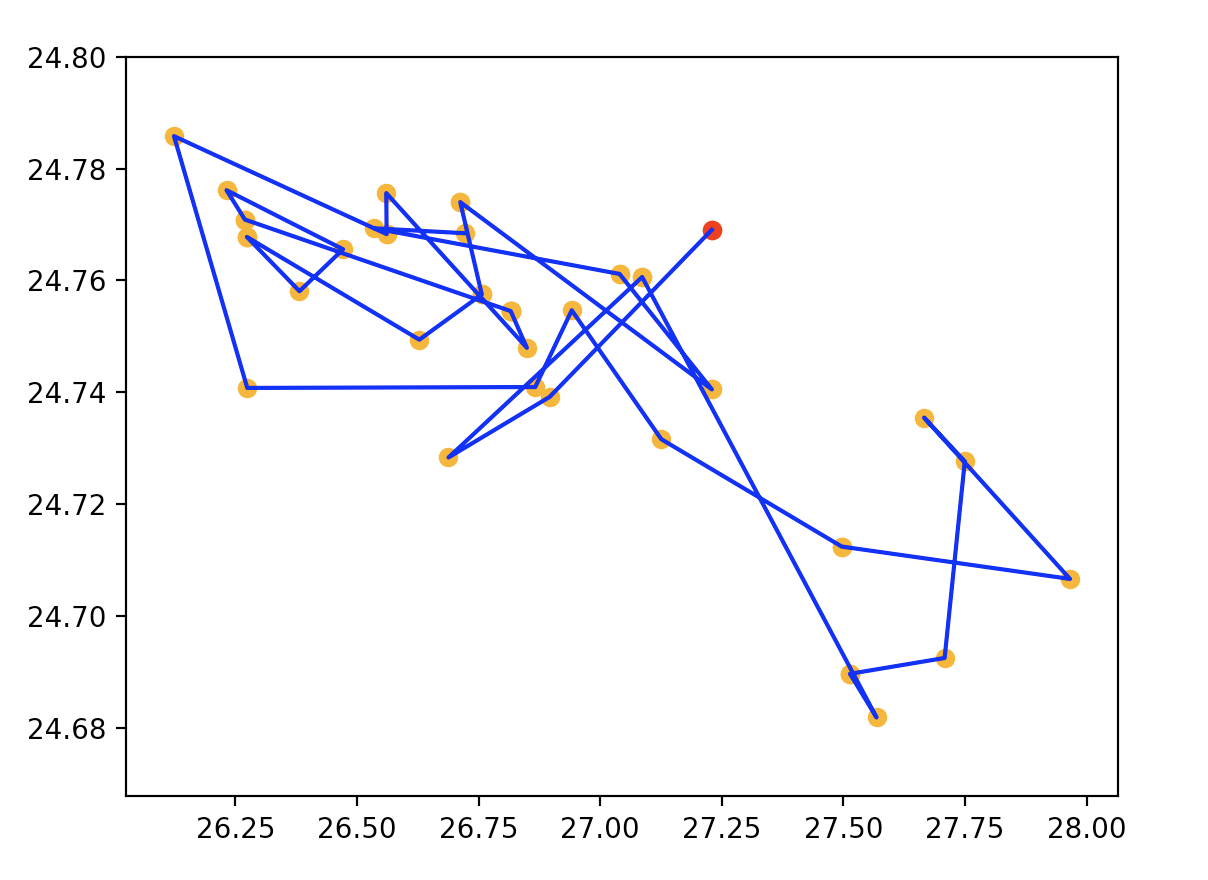}
        \caption{Walk on 5-regular graphs on 20 nodes, $p=5$}
    \end{subfigure} 
    \hspace{0.4in}
    \begin{subfigure}[b]{0.35\textwidth}
        \centering
        \includegraphics[width=\textwidth]{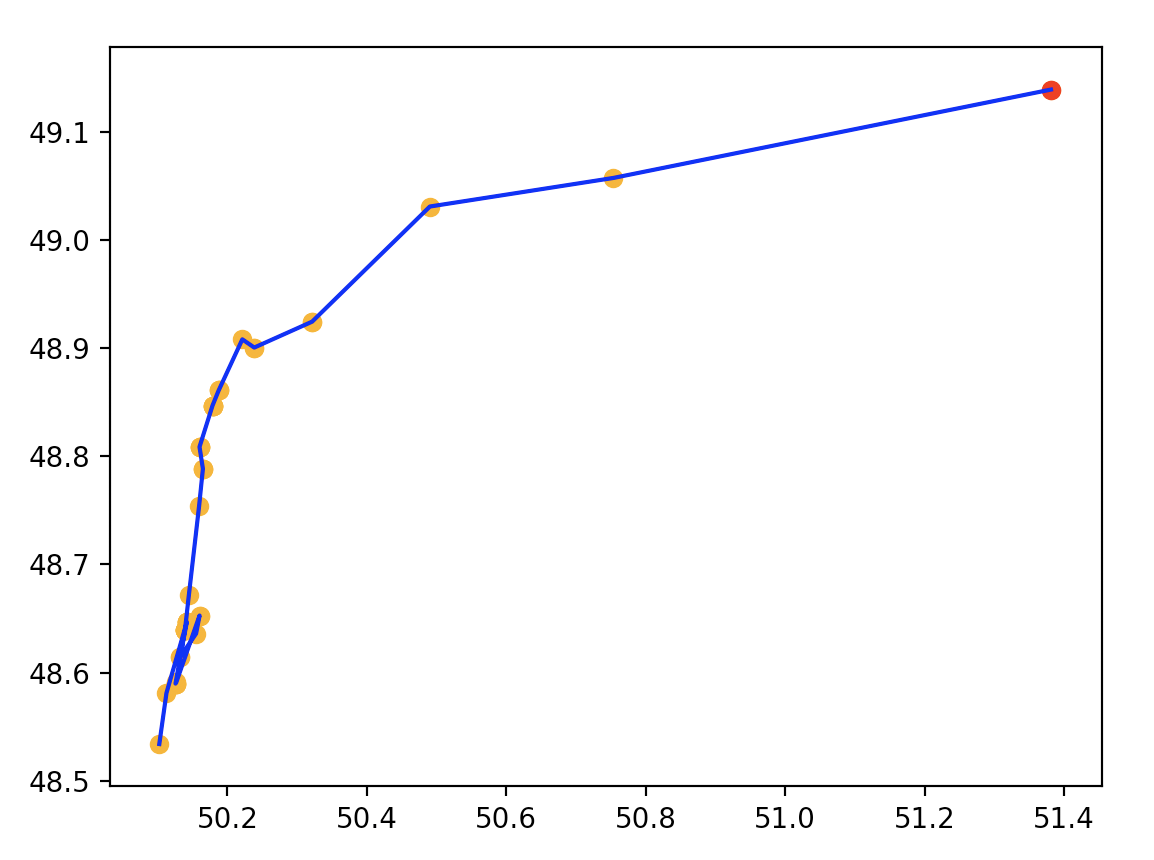}
        \caption{Walk on 8-regular graphs on 25 nodes, $p=4$}
    \end{subfigure}%
    \caption{Random walks on the set of $d$-regular graphs map to walks on the corresponding landscapes. \label{walks}
    When we are at a graph $G$, a step of the walk constitutes of randomly picking two edges with four different end points
    and replacing the edge pair randomly with one of the two other independent edge pairs on the same four nodes.
    If the move cannot be made, a new random edge pair of $G$ is picked. }
\end{figure*}

For $k=2$, we have plotted the set $\{x_{1}, x_{2}, \dots\}$ for various sets of graphs Figure \ref{landscapes}.
The pictures give a sense of how far each pair of graphs are from each other. When one looks at the landscapes of all 
three-regular graphs on 16 nodes in Figure \ref{landscapes} (we can draw as many landscapes as the number of degree sequence pairs), interestingly
one sees a structure of clusters, which we have managed to decode, and subsequently color-code. It turns out that the clusters correspond to the 
number of triangles in the graph. In fact, for three-regular graphs the depth-one landscapes 
contain only discrete points -- each point corresponds to a cluster of graphs with the same triangle number. This has lead us to realize:
\begin{theorem}[\cite{marioXx4}] The depth one QAOA values for 3-regular graphs with respect to the CUT energy function 
is solely the function of the 
degree sequence, the number of nodes and the number of triangles of the graph.
\end{theorem}

To test whether proximity of graphs coincides with ``QAOA proximity'' we have run a random walk on 3-regular graphs so that at every step we swap two independent edges for 
other two independent edges
in the manner described in 
Figure \ref{walks} (a). Then we have looked at how the QAOA values evolved under this Markov chain \cite{marioXx4}.
We realize that as the number of steps grows, the QAOA energy differences also grow Figure \ref{walks} (b).
The observation can perhaps be exploited in computational genetics, archeology or other areas,
where we face slowly changing structures representable by graphs.

\section*{Acknowledgements}
We would like to thank our colleagues from various teams in Alibaba Cloud Intelligence supporting us in the numerical experiments presented in this paper. J. C. would like to thank Shengtao Wang, Borja Peropadre for delightful discussion.

\bibliographystyle{plain}
\bibliography{main}
\end{document}